\documentclass[aps,pra,floatfix,twocolumn,groupedaddress,showpacs,lettersize,superscriptaddress]{revtex4-1}

\usepackage{hyperref}
\usepackage{amsmath}
\usepackage{amssymb}
\usepackage{color}
\usepackage{subfigure}
\usepackage{graphicx}
\usepackage[varg]{txfonts}
\usepackage{natbib}

\begin{document}

\title{\textit{In situ} imaging of vortices in Bose-Einstein condensates}

\author{Kali E. Wilson, Zachary L. Newman, Joseph D. Lowney and Brian P. Anderson}

\affiliation{College of Optical Sciences, University of Arizona, Tucson, AZ, 85721}

\date{\today}

\begin{abstract}
Laboratory observations of vortex dynamics in Bose-Einstein condensates (BECs) are essential for determination of many aspects of superfluid dynamics in these systems.  We present a novel application of dark-field imaging that enables \texttt{\it in situ} detection of two-dimensional vortex distributions in single-component BECs, a step towards real-time measurements of complex two-dimensional vortex dynamics within a single BEC.  By rotating a $^{87}$Rb BEC in a magnetic trap, we generate a triangular lattice of vortex cores in the BEC, with core diameters on the order of 400 nm and cores separated by approximately 9 $\mu$m.  We have experimentally confirmed that the positions of the vortex cores can be determined without the need for ballistic expansion of the BEC. \end{abstract}



\maketitle

Quantized vortices in superfluids serve as localized indicators of the superfluid's dynamics.  Two-dimensional (2D) vortex distributions are especially relevant in recent experimental efforts to better understand the fluid dynamics of BECs, including vortex dipole and cluster formation~\cite{Nee2010.PRL104.160401}, 2D quantum turbulence~\cite{Neely2013,Shin},  formation and decay of persistent currents~\cite{PhysRevA.86.013629,Wright2013,Neely2013,PhysRevA.89.053606}, and the Berezinskii-Kosterlitz-Thouless transition~\cite{Hadzibabic2006,Desbuquois2012,Choi2013}.  However, laboratory visualization of vortex cores in a minimally destructive manner that allows for the tracking of vortex motion has remained a considerable challenge. Given the sub-micron size of a vortex core, most experiments involving vortex imaging have relied on a period of ballistic expansion of the BEC prior to image acquisition, limiting observations to a single image of the BEC; see Ref.~\cite{Anderson2010} for an overview of such experiments.  Stroboscopic expansion and probing of small fractions of condensed atoms has enabled the determination of few-vortex dynamics in a single BEC~\cite{Fre2010.Sci329.1182}, although the utility of this technique in measurements of many-vortex dynamics is not yet known.   In order to detect the motions of numerous vortices, new imaging procedures must be explored.  Here we demonstrate single-shot \texttt{\it in situ} imaging of a 2D vortex distribution in a rotating BEC, obtained by applying a high-angle dark-field imaging technique that is similar to methods commonly employed in other applications of microscopy~\cite{Settles2001}.   With additional modifications, this imaging method should be amenable to the acquisition of multiple images of a single BEC, and hence offers the potential for experimental determination of the dynamics of 2D vortex distributions.

To date, the most versatile demonstrated method for imaging the dynamics of an arbitrary few-vortex distribution in a BEC is that of Freilich \textit{et al.}~\cite{Fre2010.Sci329.1182}, in which small fractions of the atoms from a single BEC are repeatedly extracted, ballistically expanded, and imaged. This technique allows for the acquisition of sequential absorption images of a single BEC, but since it relies on a period of expansion before vortex cores are resolvable, this method may present difficulties in determining the positions of vortices within a tightly packed vortex cluster. Additionally, the required expansion time limits the acquisition rate of these images, making the motion of more than three or four cores difficult to track as inter-vortex distances decrease and vortex core positions change more rapidly.  Minimally destructive, \texttt{\it in situ} observations of vortex dynamics in a single BEC have also been obtained by filling the vortex core with atoms in a different atomic state~\cite{And2000.PRL85.2857}. Filling the core increases the size of the vortex and enables it to be easily resolved \textit{in situ} with phase contrast imaging techniques, but interactions between the two atomic states strongly affect the dynamics of the quantum fluid.

\begin{figure}[t]
\centering\includegraphics[width=.95\columnwidth]{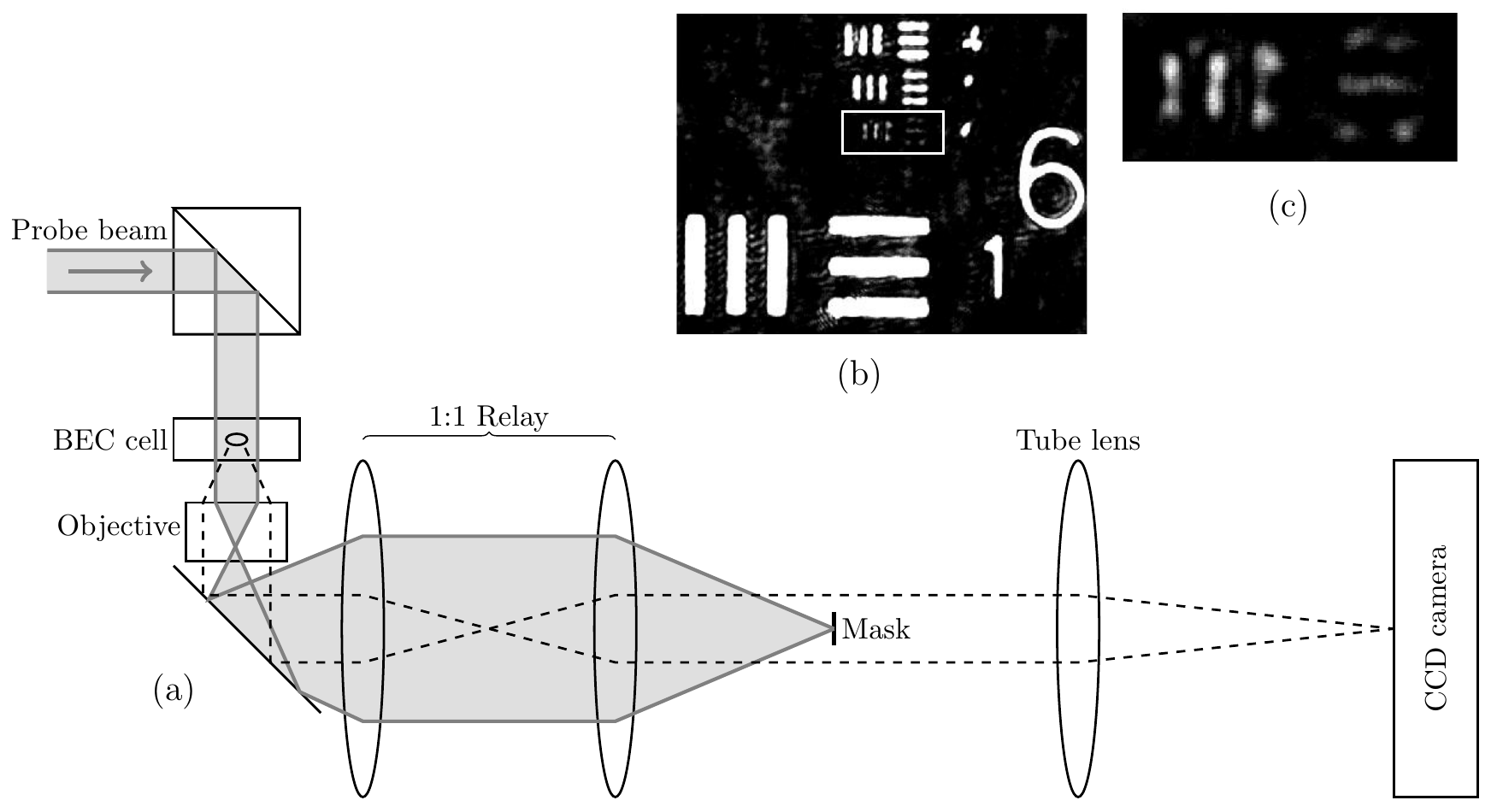}
\caption{(a) BEC imaging optics (not to scale). 780-nm probe light (shaded in gray) is directed towards the BEC along the vertical (axial) imaging axis. Light refracted by the BEC (represented by dashed lines) is collected with a microscope objective, and imaged on the CCD camera with a measured optical system magnification of M$\sim$20. A mask placed in the intermediate Fourier plane of the imaging system provides a high-pass spatial filter. (b) Image of a US Air Force resolution test target, obtained with an off-line replica of the imaging system using 780-nm laser light and showing group 6, element 1 (bottom row), and group 7,  elements 4-6 (top three rows). (c) Zoomed image of group 7, element 6, the features enclosed in the superimposed white box in (b).  These features have a line width of 2.19 $\mu$m with a center-to-center separation of 4.38 $\mu$m, setting the upper bound on the resolution of our imaging system.  The image of the target is used to determine a measured magnification of M=19.7$\pm$0.4, where the error is due to our uncertainty in measuring the periodicity in the test target image.   All images obtained with the offline imaging system were taken with a Point Grey Firefly MV CMOS camera with  6 $\mu$m x  6 $\mu$m pixels.}
\label{layout}
\end{figure}

Our imaging approach involves an adaptation of dispersive dark-field imaging~\cite{Settles2001}.  Conceptually, in dark-field imaging, the BEC is treated as a phase object that coherently refracts light from an imaging probe beam; see Ref.~\cite{Stamper-Kurn1999} for a detailed discussion of this imaging method.   Briefly, with a monochromatic probe laser beam of approximately uniform intensity $I_0$ propagating along the $z$ direction, the spatially dependent phase shift $\phi(x,y)$ acquired as the probe passes through the BEC is given by
\begin{equation}
\phi(x,y) = -\tilde{n}(x,y) ~\sigma_0 \left(\dfrac{\Delta / \Gamma}{1+4(\Delta / \Gamma)^2 + I_0/I_{\mathrm{sat}}} \right).
\end{equation}
In this expression, $\tilde{n}(x,y)=\int{n(x,y,z)~dz}$ is the $z$-integrated column density of the BEC obtained from the full atomic density distribution $n(x,y,z)$, $\sigma_0$ is the resonant atom-photon scattering cross-section, $\Delta = \omega - \omega_0$ is the detuning of the probe frequency $\omega$ from atomic resonance $\omega_0$, $\Gamma$ is the natural linewidth of the atomic transition, and $I_{\mathrm{sat}}$ is the transition saturation intensity. As illustrated in Fig.~\ref{layout}(a), an opaque mask placed on-axis in the Fourier plane of an imaging system acts as a high-pass spatial filter, blocking the unrefracted component of the probe beam, but allowing the light refracted by the BEC to reach the camera.  Andrews \textit{et al.}~\cite{Andrews05071996} first applied dispersive dark-field BEC imaging as a minimally destructive alternative to absorption imaging, and demonstrated multi-shot imaging of a single BEC. More recently, Pappa \textit{et al.}~\cite{Pappa11} employed near-resonant dark-field imaging to make highly sensitive measurements of the components of a spinor BEC, reporting a detection limit of about seven atoms.  In both of these applications of the dark-field technique, the intent was to image the bulk profile of the BEC, rather than locate microscopic features within the BEC. 

We use dark-field imaging to isolate the imaging light scattered by sub-micron features within the BEC, and in particular, to identify the positions of vortex cores.  A vortex core is free of condensed atoms, and therefore the core position corresponds to a steep density gradient over a distance on the order of the healing length~\cite{Pet2008}, approximately 400 nm for our parameters.  Such a sharply localized density feature acts as a strong lens that refracts light into high spatial frequencies. By carefully selecting the size of the dark-field mask, we remove the low spatial frequencies associated with the more gradual changes in the BEC density profile, allowing only the light refracted by the vortex cores to reach the camera. Without the large background signal of the bulk BEC, it is then feasible to pick out the refracted signal due to each vortex core without expanding the BEC.   We describe this process as \textit{in situ} vortex imaging due to the ability to detect vortex cores without using a period of ballistic expansion, although all BEC imaging procedures are at least somewhat destructive.

For the images of vortices reported here, we formed BECs of $5\,^2S_{1/2}$ $|\mathrm{F} = 1, m_{\mathrm{F}} = -1\rangle$ $^{87}$Rb atoms in a magnetic time-averaged orbiting potential (TOP) trap~\cite{PhysRevLett.74.3352}, with radial and axial trap frequencies of $(\omega_{\mathrm{r}}$, $\omega_{\mathrm{z}}) \sim 2\pi \times (8,16) ~\mathrm{Hz}$, BEC atom numbers of approximately 1.8 x 10$^6$, and BEC Thomas-Fermi radii of $(R_{\mathrm{r}}, R_{\mathrm{z}}) \sim (35, 19) ~\mu\mathrm{m}$. Following Hodby \textit{et al.}~\cite{Hod2001.PRL88.010405}, we modified the TOP trap's rotating bias field to form a slowly rotating elliptical potential well, which in turn spun up the BECs such that a triangular lattice of vortices was formed.  The vortex lattice provided a reproducible and easily recognizable pattern of vortex cores for our imaging tests.

As illustrated in Fig.~\ref{layout}, our imaging system consists of an infinite-conjugate Olympus SLMPLN 20X microscope objective with a numerical aperture NA=0.25, a theoretical diffraction-limited resolution of 1.9 $\mu$m at a wavelength of $\lambda$=780 nm \cite{note2}, a working distance of 25 mm, and a focal length of 9 mm.  The objective is followed by a 1:1 relay lens pair, comprised of two 75-mm focal length achromatic doublets separated by 150 mm.  The dark-field imaging mask is placed at the intermediate Fourier plane, located at the back focal plane of the relay, between the final relay lens and the tube lens.  The relay lens pair is necessary because the initial Fourier plane where the mask would ideally be placed is located within the objective lens housing.  Finally, a singlet lens with a focal length of 175 mm is used as the tube lens. All BEC images were obtained with a Princeton Instruments PIXIS 1024 BR back-illuminated CCD camera with 13 $\mu$m x 13 $\mu$m pixels. The imaging system has a magnification of M=19.7$\pm$0.4.  We used a variety of dark-field masks and sizes in our imaging tests, described below.

\begin{figure}[t]
\centering\includegraphics[width=\columnwidth]{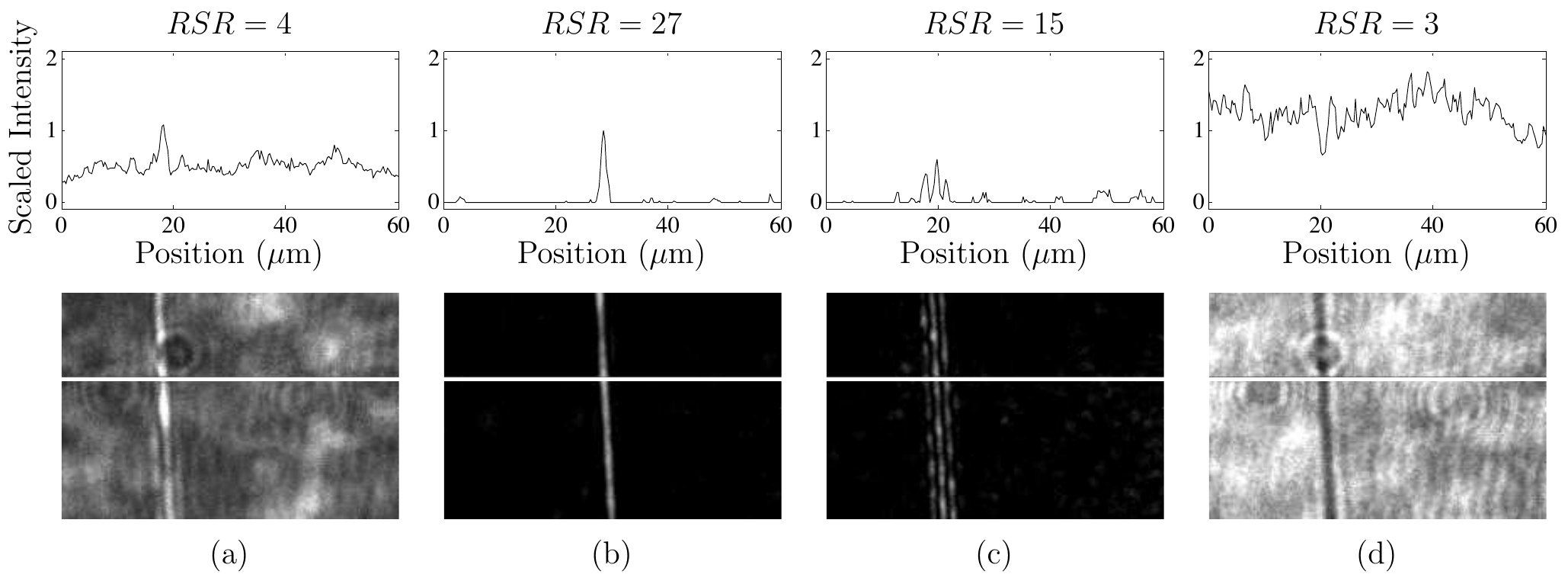}
\caption{Raw 60-$\mu$m-wide images (lower panels) of a section of silica nanofiber with a diameter of $\sim$500 nm (vertically oriented in each image in the lower panels), shown without the use of background subtraction or other signal-enhancing techniques.  660-nm imaging light was used for all images. Upper panels show horizontal cross sections through each corresponding lower image along the white line superimposed on the images; image intensity is plotted (arbitrary units are the same for each cross section). (a) - (c) Dark-field images taken with a 2.5-ms exposure and using masks with diameters of 100 $\mu$m, 370 $\mu$m, and $\sim$1.5 mm respectively. Circular masks were used for (a) and (c), whereas a wire mask, aligned approximately parallel to the fiber, was used for (b). (d) Bright-field image of the nanofiber with no mask in place, taken with a 0.25-ms exposure.  See text for a discussion of the calculated relative signal ratio (RSR) for the cross sections \cite{note1}. }  
\label{DG_nano}
\end{figure}

As a first test of the capabilities of our imaging system, we constructed the system off-line with a microscope slide in place of one of the 1-mm-thick glass cell walls of the vacuum chamber, and imaged a silica nanofiber~\cite{Stiebeiner10,Vetsch10} with 660-nm imaging light. The nanofiber, with a diameter of approximately 500 nm, provides an example of a sub-micron phase object with approximately the same diameter as a vortex core, and thus serves as a suitable imaging test object.   Figures \ref{DG_nano}(a)-(c) show images of the nanofiber obtained by varying dark-field mask size, each image acquired using a 2.5-ms exposure time. The bright-field image shown in Fig.~\ref{DG_nano}(d) is provided for comparison.

Despite its sub-micron diameter, the nanofiber's position can be clearly determined in both the bright-field and dark-field images. However, in the bright-field image shown in Fig.~\ref{DG_nano}(d), the detected signal depth from the nanofiber is the same order of magnitude as background features due to structure on the probe beam, with a relative signal ratio (RSR) of 3, a measure of the relative height of the signal compared with the variations and noise in the background signal~\cite{note1}. As shown in Figs.~\ref{DG_nano}(a)-(c), RSR increases with mask size until the mask begins to alter the profile of the nanofiber signal due to clipping of the lowest spatial-frequency components in the refracted signal.  The dark-field image of Fig.~\ref{DG_nano}(b) has a RSR of 27, almost an order-of-magnitude improvement over the bright-field image.  The 370-$\mu$m-diameter wire mask used for the image of Fig.~\ref{DG_nano}(b) is the same wire used for \textit{in situ} vortex imaging described below, and the resulting image of the nanofiber has a full width at half maximum (FWHM) of 1.18 $\pm$ 0.03 $\mu$m, a measure of the lower bound of the resolution of the imaging system rather than a measure of the true size of the nanofiber. The FWHM was found by fitting a Gaussian to the intensity profile, and the uncertainty is due to the uncertainty from the fit combined with the uncertainty reported above for the system magnification. For the 660-nm probe wavelength, the calculated diffraction limit of the objective is 1.61 $\mu$m \cite{note2}, which corresponds to a FWHM of 1.36 $\mu$m for a diffraction-limited point object. Note that the high-pass spatial-frequency filtering inherent in the dark-field imaging process acts to narrow the FWHM while increasing the intensity in the side lobes of the Airy diffraction pattern.  Because of this filtering process, it is possible to obtain an image of a point object that has a FWHM that is slightly smaller than the diffraction limit, as we observe.

Although our ultimate goal is to image arbitrary 2D vortex distributions in highly oblate BECs, we chose a vortex lattice for our initial \textit{in situ} imaging tests because a lattice is an easily recognizable pattern of vortices that can be reliably reproduced. Additionally, the increase in angular momentum due to rotating the BEC causes the BEC's radial width to increase, its axial width to decrease, and the vortices comprising the lattice to align with the rotation and imaging axis. A rotating BEC thus serves as a suitable proof-of-principle test for investigating the possibility of imaging arbitrary 2D vortex distributions in highly oblate BECs, which we ultimately intend to study.

\begin{figure}[t]
\centering\includegraphics[width=.95\columnwidth]{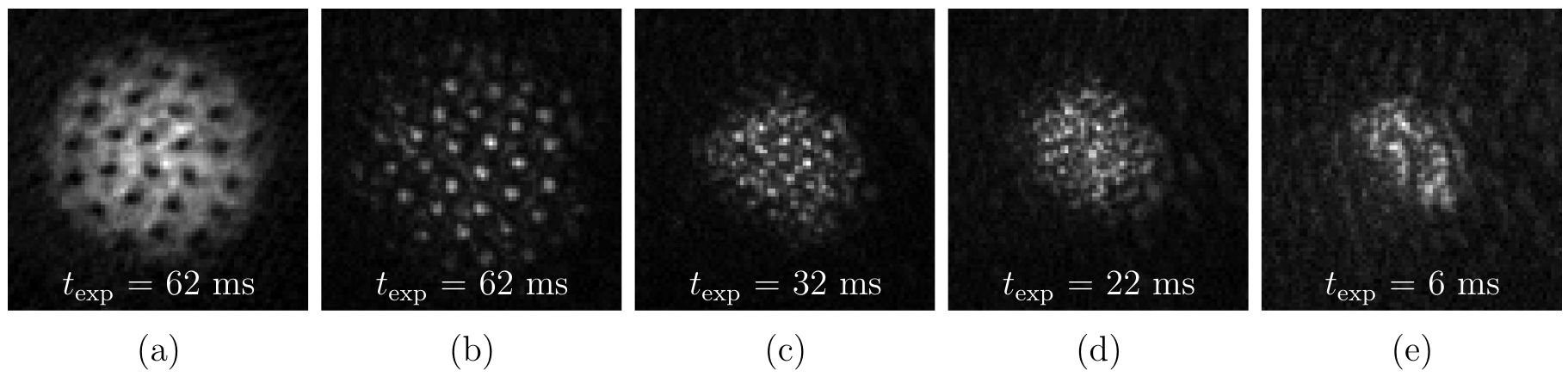}
\caption{200-$\mu$m-wide images of BECs with a vortex lattice, using an M = 5, NA = 0.2 imaging system. For each image, the BEC was released from the trap and allowed to expand for a variable time $t_{\mathrm{exp}}$ shown on each image. (a) Reference absorption image of an expanded BEC with a vortex lattice, obtained using standard methods of bright-field imaging with background subtraction and grayscale contrast inversion.   (b)-(e) Raw dark-field images of an expanded BEC with a vortex lattice taken at varying expansion times, with no background subtraction. A circular mask with diameter of $\sim$ 1.6 mm was used for all dark-field images. The lattice becomes unresolvable between $t_{\mathrm{exp}}$ = 32 ms and $t_{\mathrm{exp}}$ = 22 ms. }
\label{DG_5x}
\end{figure}

As a basis for comparison of the new optical system's imaging abilities to measure vortex distributions, we used our standard, non-diffraction-limited, M = 5, NA = 0.2 imaging system to obtain dark-field images of vortices with a BEC after a period of expansion. After spinning up a lattice, turning off the trapping fields, and allowing the BEC to expand for 62 ms,  we optically pumped the atoms from the $5\,^2S_{1/2}$ $|\mathrm{F} = 1\rangle$ level to the  $|\mathrm{F} = 2\rangle$ level and then imaged on the transition to the $5\,^2P_{3/2}$ $|\mathrm{F}' = 3\rangle$ level.  We obtained images of vortex cores using both standard bright-field absorption imaging, as shown in Fig.~\ref{DG_5x}(a), and dark-field imaging, as shown in Fig.~\ref{DG_5x}(b). A circular mask with a diameter of $\sim$1.6 mm was used for all of the dark-field images shown in Fig.~\ref{DG_5x}, and the probe detuning ranged from -1$\Gamma$ to -2$\Gamma$ from the $|\mathrm{F} = 2\rangle$ to $|\mathrm{F}' = 3\rangle$ hyperfine transition. As shown in Figs.~\ref{DG_5x}(b)-(e), vortex core resolvability decreased for shorter expansion times. The low magnification and NA of the imaging system limited our ability to resolve two neighboring cores for expansion times less than about 30 ms. Additionally, a shorter expansion time, with the corresponding smaller atom cloud, should result in a higher percentage of the light refracted from the bulk BEC bypassing the mask, thereby reducing contrast between vortex cores and the bulk BEC.

\begin{figure}[t]
\centering\includegraphics[width=.9\columnwidth]{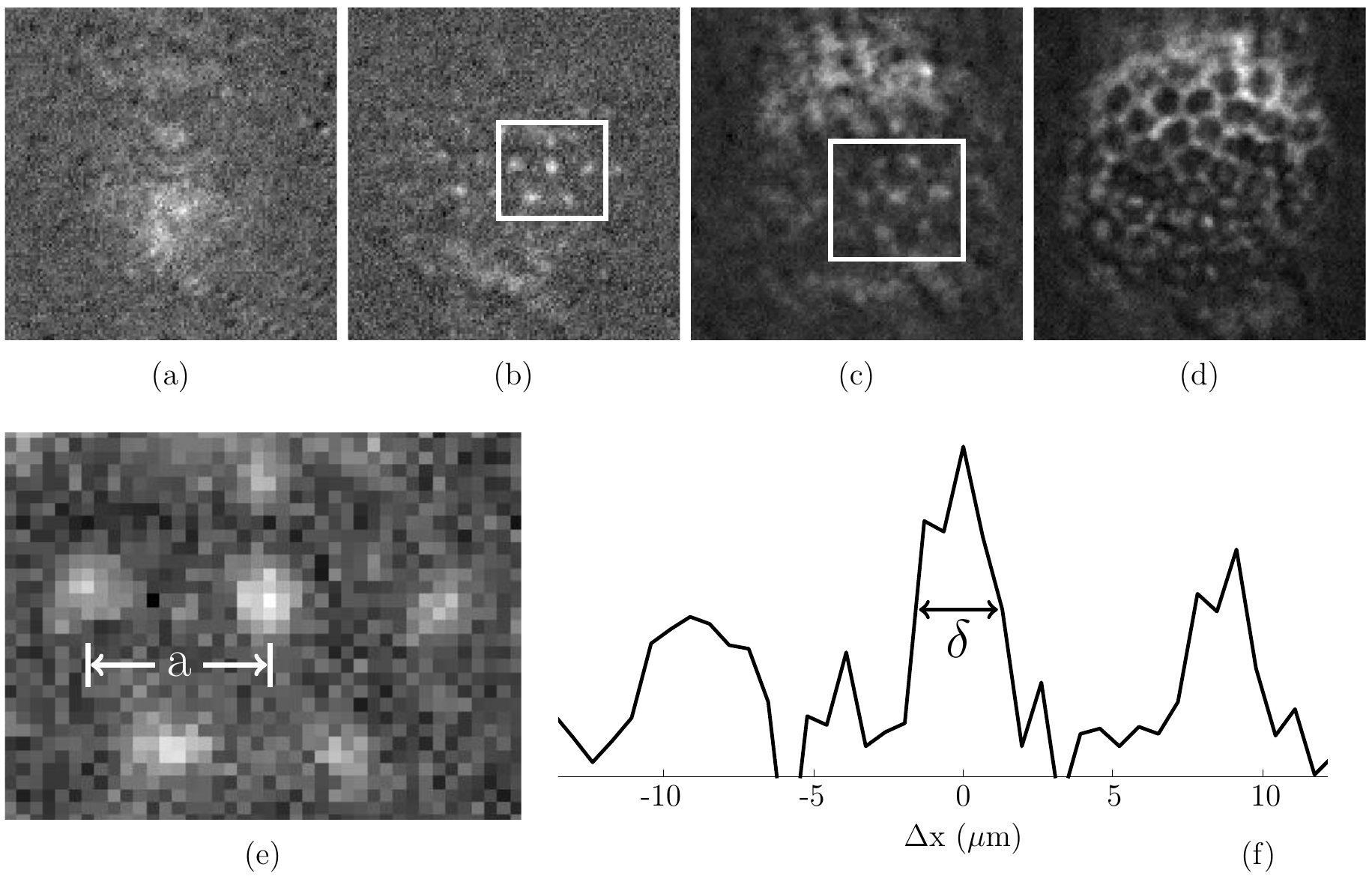}
\caption{80-$\mu$m-wide \texttt{\it in situ} images of BECs obtained with our M$\sim$20, NA = 0.25 imaging system. Images have been processed by subtraction of a background image taken in the absence of a BEC. Dark-field images of the BEC are shown, without a vortex lattice (a), and with a lattice (b). The wire mask used for both images had diameter of 370 $\mu$m, and was aligned horizontally with respect to the image. (c) Dark-field image of a BEC with a vortex lattice, but with a 250-$\mu$m-diameter wire mask; vortices are visible within the superimposed white rectangle, but not as apparent as in (b), and additional refracted light reaches the camera.  (d) Dark-field image obtained with a rotating BEC with a vortex lattice when the imaging system is not properly focused.  (e) Zoomed view of the region bounded by the white rectangle in (b), with pixelation due to the 13~$\mu$m x 13~$\mu$m camera pixels. Neighboring vortex cores are separated by $a \sim$ 9~$\mu$m. (f) Cross-section along the middle row of vortex cores shown in (e); vertical scale is proportional to pixel intensity; horizontal axis shows the distance $\Delta x$ away from the central vortex core, with the scale corresponding to real distances at the object plane.  The FWHM of the central vortex core is $\delta$=2.4$\pm$0.5~$\mu$m. The The FWHM was found by fitting a Gaussian to the intensity profile, and the reported error is due to the uncertainty from the fit. }
\label{DG_20x}
\end{figure}

To image vortex cores within a trapped BEC, we used the M$\sim$20, NA = 0.25 imaging system previously described. Representative \textit{in situ} dark-field images of a BEC confined within the TOP trap are shown in Fig.~\ref{DG_20x}. The dark-field images show a clear distinction between a BEC without a vortex lattice, Fig.~\ref{DG_20x}(a), and one with a lattice, Figs.~\ref{DG_20x}(b)-(e).  For images  Fig.~\ref{DG_20x}(a)-(c), we used an imaging probe with a $1/e^2$ beam radius of $\sim$ 2 mm, a power of $\sim$ 0.5 mW, detuning of $\Delta=4.5\Gamma$ from the $|\mathrm{F} = 2\rangle$ to $|\mathrm{F}' = 3\rangle$ transition, and an exposure time of 20 $\mu$s. For Fig.~\ref{DG_20x}(b), we measured the separation between vortex cores to be $a \sim$ 9 $\mu$m, as shown in  Fig.~\ref{DG_20x}(e).  The FWHM of the central vortex core, shown in Fig.~\ref{DG_20x}(f), was measured to be $\delta$=2.4$\pm$0.5 $\mu$m, indicating that the imaging system should be capable of resolving two vortex cores separated by approximately this distance. The FWHM is a measure of the point-spread function for our imaging system rather than the actual size of the vortex core. The detuning was chosen to maximize the signal from the vortex cores for the 370-$\mu$m-diameter mask.  Such close detuning was destructive to the BEC, and with these parameters we are limited to acquiring a single image per BEC. Additionally, due to the low signal level of these images, we utilized background subtraction to remove features due to the un-refracted probe light that were not obstructed by the mask.   

As discussed, the size of the dark-field mask determines the spatial frequency cutoff of the spatial filter. Figure \ref{DG_20x}(c) shows dark-field images taken using a mask diameter of 250 $\mu$m. The smaller mask size allows more of the light refracted from the bulk BEC to reach the camera, reducing the contrast of the vortex cores.  In comparison, the 370-$\mu$m-diameter wire mask used for Fig.~\ref{DG_20x}(b) blocks almost all of the light refracted by the bulk BEC. Fig.~\ref{DG_20x}(d) shows a representative out-of-focus image of a vortex lattice, obtained with a detuning of $\Delta=4\Gamma$ and a 30-$\mu$s exposure. The lattice takes on a honeycomb appearance similar to that observed in an out-of-focus bright-field absorption image of a vortex lattice. Note that with the exception of the out-of-focus lattice, we primarily see vortex cores in the center of the BEC. We speculate that this could be due to the decrease in density at the edge of the BEC, and correspondingly smaller angles of refraction due to the increase in healing length.  Additionally, the use of a wire mask introduces an asymmetry in the background signal since all spatial frequencies are blocked in the direction parallel to the wire. We anticipate that using a precision circular mask with an optimized size, and a BEC held in a flat-bottomed potential~\cite{PhysRevLett.110.200406}, will improve our ability to detect vortices across the BEC.   

The single-shot, \textit{in situ} images of bare vortex cores presented here serve as a promising proof-of-principle indication that complex vortex dynamics can be measured in a trapped BEC with additional optimization of the imaging system and imaging parameters.  In numerical studies of 2D quantum turbulence, our particular area of interest, vortex-antivortex annihilation and bound pairs of vortices of the same sign of circulation appear to show minimum inter-vortex separation distances of approximately $\sim$ 2 $\mu$m for our parameters~\cite{Neely2013}.  Resolutions approaching this scale are already achievable with our imaging system.  

The primary hurdle in extending this technique to capturing multiple images of a single BEC is the achievement of a sufficient RSR, given the large probe detuning and low probe intensity necessary for minimally-destructive imaging.  One significant advantage of dark-field imaging is the minimization of background light, allowing for weak signals to be obtained and amplified without the need for background image subtraction. This potential advantage will be especially useful for measurements of vortex dynamics where the time between images is expected to be on the order of 10 ms. In the dark-field \texttt{\it in situ} images presented here, background image subtraction was necessary due to low signal levels and relatively high levels of weakly scattered probe light reaching the camera, and further optimization of the probe beam profile and dark-field mask will be necessary to utilize raw images without the need for background subtraction.

While the Olympus objective used for our imaging system appears to be a suitable commercial objective given the physical constraints of our apparatus, this microscope objective is optimized for visible light, and its transmission is approximately 60$\%$ for our operating wavelength of 780 nm. Additionally, the relay lenses required to place the mask in an accessible intermediate Fourier plane introduce aberrations to the imaging system, making it more difficult to block all of the weakly scattered imaging light. 

In addition to optimizing the imaging parameters mentioned above, we are currently implementing other modifications that should improve both the image quality and the RSR. We are installing a custom objective, optimized for 780-nm imaging probe light, with an accessible back focal plane, based on the design of Ref.~\cite{Alt2002142}.  We also anticipate that using a CCD camera with electron multiplying (EMCCD) gain capabilities, in conjunction with dark-field imaging, will result in a significant increase in the overall signal-to-noise ratio, and will enable the use of imaging light further detuned from resonance. Recently Gajdacz \texttt{\it et al.}~have used an EMCCD camera and dark-field Faraday imaging to obtain thousands of images of a single BEC~\cite{Arlt13}.  In situations with low signal, but also low background light levels, the pre-readout amplification of an EMCCD camera should be beneficial in imaging vortex distributions.

We have demonstrated single-shot \textit{in situ} imaging of vortex cores in a BEC. Based on this result, we anticipate that improvements will enable minimally destructive, multi-shot, \textit{in situ} imaging of vortices and their dynamics within a single BEC.  Access to such images will open up new possibilities for experiments to study numerical and theoretical predictions of 2D quantum turbulence~\cite{Bradley2012,White2012a,Reeves2013,Billam2013}, our primary goal, and an even wider range of superfluid dynamics.
  
\begin{acknowledgments}
We acknowledge the support of the US National Science Foundation grant PHY-1205713.  KEW acknowledges support from the Department of Energy Office of Science Graduate Fellowship Program, administered by ORISE-ORAU under contract no.~DE-AC05-06OR23100. KEW and ZLN acknowledge support from the University of Arizona TRIF Program. We thank Olympus for the loan of the microscope objective, Pascal Mickelson and Poul Jessen for providing the nanofiber, and Tom Milster for helpful suggestions.
\end{acknowledgments}


\begin{thebibliography}{28}
\bibitem [{not()}]{note2}%
  \BibitemOpen
  \href@noop {} {}\bibinfo {note} {The diffraction-limited resolution $R$ of the microscope objective is calculated using the Rayleigh criterion, $R = 0.61\lambda/NA$.}\BibitemShut {Stop}%
\bibitem [{not()}]{note1}%
  \BibitemOpen
  \href@noop {} {}\bibinfo {note} {$RSR$ is the average of all the
  $RSR_{\mathrm{cross}}$ values calculated for individual cross-sections
  perpendicular to the nanofiber orientation. $RSR_{\mathrm{cross}} =
  |I_{\mathrm{sig}} - \bar{I}_{\mathrm{BG}}| / \sigma_{\mathrm{BG}}$,
  where $I_{\mathrm{sig}}$ is the pixel value at the peak of the nanofiber
  signal, $\bar{I}_{\mathrm{BG}}$ is the average pixel value, and
  $\sigma_{\mathrm{BG}}$ is the standard deviation. When calculating
  $\bar{I}_{\mathrm{BG}}$ and $\sigma_{\mathrm{BG}}$, the 14 pixels on
  either side of the signal peak are not used.}\BibitemShut {Stop}%
\end{thebibliography}
\end{document}